\documentclass[preprint,preprintnumbers,amsmath,amssymb]{revtex4}

\usepackage{graphicx}
\usepackage{dcolumn}
\usepackage{bm}
\usepackage{psfrag}
\usepackage{color}
\usepackage{bbm}

\usepackage[latin1]{inputenc}
\usepackage[T1]{fontenc}

\begin{document}

\preprint{UCL-IPT-05-11}

\title{A New Look at an Old Mass Relation}

\author{J.-M. G\'erard}
\email[]{gerard@fyma.ucl.ac.be}
\affiliation{Unit\'e de Physique Th\'eorique et Math\'ematique\\
Universit\'e Catholique de Louvain}

\author{F. Goffinet}
\email[]{goffinet@fyma.ucl.ac.be}
\affiliation{Unit\'e de Physique Th\'eorique et Math\'ematique\\
Universit\'e Catholique de Louvain}

\author{M. Herquet}
\email[]{mherquet@fyma.ucl.ac.be}
\affiliation{Unit\'e de Physique Th\'eorique et Math\'ematique\\
Universit\'e Catholique de Louvain}

\date{October 21, 2005}

\begin{abstract}
New data from neutrino oscillation experiments motivate us to extend a successful mass relation for the charged leptons to the other fundamental fermions. This new universal relation requires a Dirac mass around $3\ 10^{-2}\,  \mathrm{eV}$ for the lightest neutrino  and rules out a maximal atmospheric mixing. It also suggests a specific decomposition of the CKM mixing matrix. 
\end{abstract}

\maketitle

\section{\label{sec:Intro}``Who ordered that ?''}
The spectrum of lepton and quark masses puzzles particle physicists since the muon discovery in cosmic rays.  Today, only the heaviest fermion appears to have a rather natural mass in the electroweak unification theory at the Fermi scale, but the leptonic mass ratio
\begin{equation*}
\frac{m_\mu}{m_e}\sim 200
\end{equation*}
remains totally unexplained in this framework.

The empirical mass relation for the charged leptons \cite{Koide:1982si,Koide:1989jq}
\begin{equation}
m_{e}+m_{\mu}+m_{\tau}=\frac{2}{3}(\sqrt{m_{e}}+\sqrt{m_{\mu}}+\sqrt {m_{\tau}})^2
\label{eq:koide}
\end{equation}
is now precise at the level of $10^{-5}$ : if we switch off the electron mass, we obtain a completely wrong tau-to-muon mass ratio.  In that sense, Rabi's famous question about the muon can be rephrased in a milder way since
\begin{equation*}
  \frac{m_\tau}{m_\mu}\sim 20\ .
\end{equation*}
Yet, at first sight this remarkable but mysterious mass relation seems to be a false trail.  Any attempt to apply \eqref{eq:koide} to the quarks is indeed doomed to failure.  The down quark family might fulfil this simple non-linear mass relation since
\begin{equation*}
  \frac{m_b}{m_s}\sim \frac{m_\tau}{m_\mu} 
\end{equation*}  
with large theoretical uncertainties on the value of $m_s$. However, the up quark family with
\begin{equation*}
  \frac{m_t}{m_c}\sim \frac{m_\mu}{m_e} 
\end{equation*}
revives Rabi's worry and removes any hope to have a universal fermion mass relation at our disposal, whatever the value $m_u$ may be.

Now, let us have a closer look at the democratic relation \eqref{eq:koide}.  The challenging middle value of 
\begin{equation}
\label{eq:q}
q \equiv \frac{\sum m_i}{(\sum \sqrt{m_i})^2}
\end{equation}
turns out to be an extremely efficient measure of the mass splitting inside the charged lepton family.  Indeed, $q^e = \frac{2}{3}$ together with the physical electron and muon masses predicted $m_{\tau} = 1776.97 \ \text{MeV}$ before precise measurements.  Its maximal value ($q=1$) would correspond to a full hierarchy ($m_1 \ll m_2 \ll m_3$) while its minimal value ($q=1/3$) should occur in the case of complete degeneracy ($m_1 = m_2 = m_3$). Nature comes close to these boundary values with the up quark and the neutrino families, respectively.

We argue that the flavour mixing which also displays quite different patterns, from the small angles of the Cabibbo-Kobayashi-Maskawa (CKM) matrix for the quarks to the large ones of the Maki-Nakagawa-Sakata (MNS) matrix for the leptons, holds the key of this puzzle.  In models with flavour symmetries, small angles are closely linked to mass hierarchy and large ones to mass degeneracy. This points us the way towards a universal mass/angle relation.

In this Letter, we take advantage of the recent experimental progress in the neutrino sector to generalize the mass relation \eqref{eq:koide} for the lepton families. A lepton-quark connection beyond the Standard Model is then called upon to validate it also for the quark families.


\section{Lifting the near degeneracy in the neutrino family}
It is now an experimental fact that the neutrinos are massive and mix. However, if their mixing is relatively well measured, their mass scale is still unknown and an alternative remains for the hierarchy, either normal ($m_1 < m_2 \ll m_3$) or inverted ($m_3 \ll m_1 < m_2$).  The latest results at $1\sigma$ are \cite{Strumia:2005tc}
\begin{eqnarray*}
\Delta m^2_{21} =& m^2_2 - m^2_1 &= (8.0 \pm 0.3) \times 10^{-5} \ \mathrm{eV}^2\\
| \Delta m^2_{32}| =& |m^2_3 - m^2_2| & = (2.5 \pm 0.3) \times 10^{-3} \ \mathrm{eV}^2 \quad.
\end{eqnarray*}

\enlargethispage*{1cm}

A first attempt to apply the original mass relation \eqref{eq:koide} to the neutrinos was unsuccessful in both schemes \cite{Li:2005rp}.  The reason simply lies in the mild splitting of the neutrino masses: the strongest hierarchy, ensured  for $m_1 = 0$, always implies $q^\nu < 0.6$.  One way out is to amplify the mass hierarchy with the help of the well-measured neutrino mixing matrix elements $(U_L^{\nu})^{ij}$.  We thus propose the following minimal extension of relation \eqref{eq:koide}:
\begin{equation}
\label{eq:new}
\sum \tilde{m}_i = \frac{2}{3} \left( \sum \sqrt{\tilde{m}_i}\right)^2
\end{equation}
which acts on the ``pseudo-masses'' $\tilde{m}_i$ defined as  
\begin{equation}
\label{eq:WBmass}
  \tilde{m}_i\equiv  | \sum_j U_{L}^{ij}\  m_j  | \quad,
\end{equation}
rather than on the physical masses $m_j$. In our convention, $U_L^\dagger \ M U_R \equiv \mathrm{diag}(m_1,m_2,m_3)$ such that if $U_R=\mathbbm{1}$, these Dirac pseudo-masses are simply related to the Yukawa couplings of a single Higgs doublet.

The latest results from neutrino experiments  at $1\sigma$ \cite{Strumia:2005tc} are in good agreement with $\theta_{13}=0$: $$\sin^2 2 \theta_{13} = 0 \pm 0.05 \quad .$$   So, let us therefore assume the following MNS matrix
\begin{equation}
  \label{eq:mns}
  V_{MNS}\equiv U_L^{e^\dag} U_L^{\nu}= R_{23}(\theta_\oplus)R_{12}^T(\theta_\odot)=
\begin{pmatrix}
1 & 0 & 0 \\
0 & \cos \theta_\oplus & \sin \theta_\oplus \\
0 & -\sin \theta_\oplus  & \cos \theta_\oplus  \\
\end{pmatrix}
\begin{pmatrix}
\cos \theta_\odot & -\sin \theta_\odot & 0 \\
\sin \theta_\odot & \cos \theta_\odot  & 0 \\
0 & 0 & 1 \\
\end{pmatrix}
\end{equation}
with the experimental values for the mixing angles at 1$\sigma$ \cite{Strumia:2005tc}
\begin{align*}
  \label{eq:neutrinodata}
  \tan^2 \theta_\odot &= 0.45\pm 0.05\\
  \sin^2 2\theta_\oplus &= 1.02\pm 0.04 \quad.
\end{align*}

There are three natural solutions in this limit \cite{Barr:2002qz}.   Either the large solar mixing angle $\theta_\odot$ comes from $M^\nu$ and the large atmospheric angle $\theta_\oplus$ from $M^e$, or both come from $M^e$ or $M^\nu$.   The remarkable accuracy of the relation \eqref{eq:koide} requires that any successful extension involving mixing angles should reduce to this form for the charged leptons.  Consequently, we  focus on the last possibility, namely
\begin{equation}
  \label{eq:Unu}
  U_L^{e}=\mathbbm{1}\quad,\quad U_L^{\nu}=R_{23}(\theta_\oplus)R_{12}^T(\theta_\odot)\quad.
\end{equation}

\begin{figure}[t!]
\psfrag{m12}{$\Delta m_{21}^2 (10^{-5} \text{eV}^2)$}
\psfrag{m23}{$\Delta m_{32}^2 (10^{-3} \text{eV}^2)$}
\psfrag{tan2th12}{$\tan^2 \theta_\odot$}
\psfrag{sin2th23}{$\sin^2 \theta_\oplus$}
\includegraphics[scale=0.52]{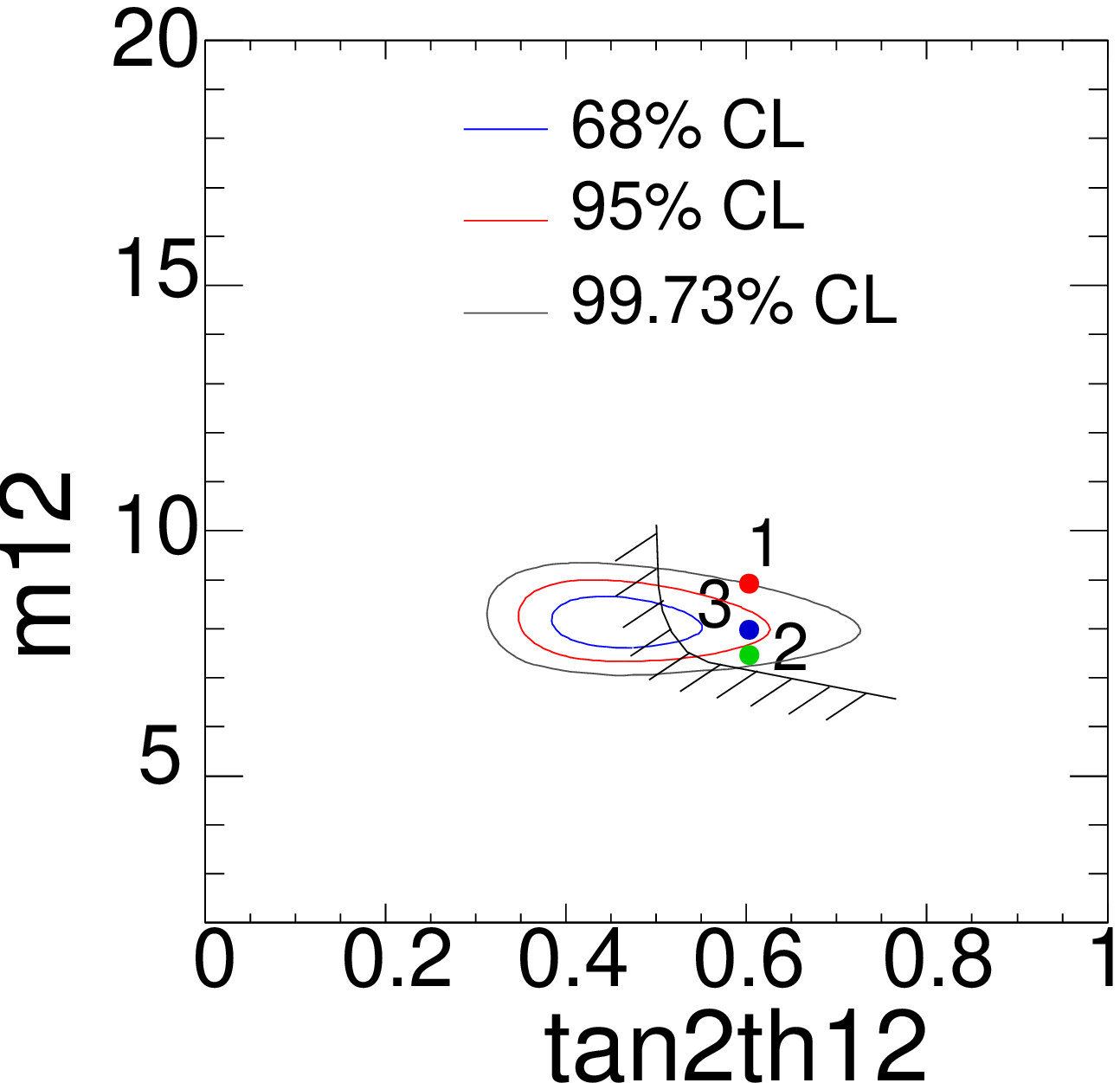}
$\quad$
\includegraphics[scale=0.83]{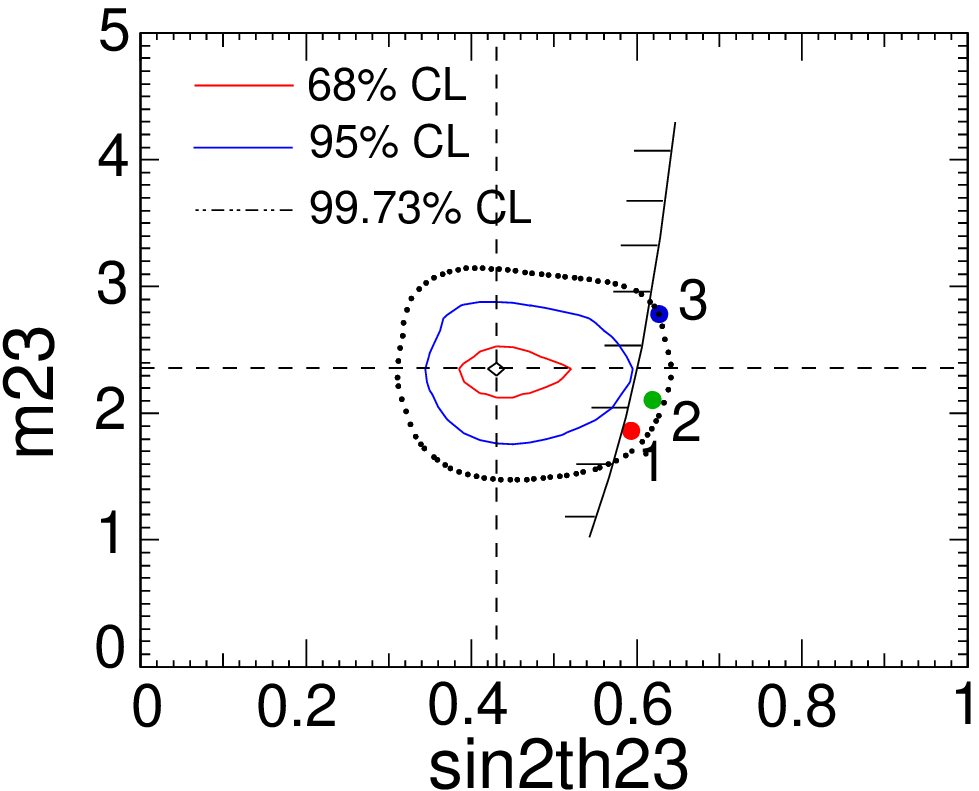}
\caption{\label{fig:neutrino}Exclusion domains and three ``sample'' points derived from relation \eqref{eq:new}. The solar and reactor experimental results have been taken from the latest global fit \cite{Poon:2005} while the atmospheric neutrino data come from the combined SK and K2K results \cite{Fogli:2005gs}.}
\end{figure}

Numerical computations provide us with a continuous set of solutions satisfying \eqref{eq:new} and compatible with the present data (see Fig.~\ref{fig:neutrino}).  All these solutions correspond to the normal hierarchy for $\lbrace m_i \rbrace$, the inverted one being excluded.  For illustration, three typical solutions are also displayed on Figure \ref{fig:neutrino} and the corresponding numerical results are listed in Table \ref{tab:neutrino}.
It appears that the predicted range for the Dirac mass of the lightest neutrino  $m_1$ is
\begin{equation}
2\;10^{-2}\text{ eV}<m_1<4\;10^{-2}\text{ eV}
\end{equation}
at 99\% CL.  Notice that the normal hierarchy for $\lbrace m_i \rbrace$ is turned into an inverted one for $\lbrace \tilde m_i \rbrace$, i.e. $0 \approx \tilde m_3 \ll \tilde m_1 < \tilde m_2$. On the other hand, the solar $\theta_\odot$ and the atmospheric $\theta_\oplus$ angles are bounded from below by  
\begin{equation}
\begin{aligned}
\theta_\odot&> 35^\circ \\
\theta_\oplus&> 50^\circ 
\end{aligned}
\end{equation}
such that the so-called maximal mixing solution ($\theta_\oplus = \pi/4$) is ruled out.  
\definecolor{mygreen}{rgb}{0,0.5,0}
\begin{table}[htbp]
  \centering
  \begin{tabular}{|c|c|c|c|c|c|}
    \hline
    \small{Solution} &$\theta_\odot$ &$\Delta m_{21}^2$ ($10^{-5} \text{eV}^2$)& $\theta_\oplus$ &$\Delta m_{32}^2$ ($10^{-3} \text{eV}^2$)&$m_1$ ($10^{-2} \text{eV}$)\\
    \hline
    \hline
    \color{red}{1}& $37.44^\circ$ & 8.7 &$50.30^\circ$ & 1.80 & 3.14\\
    \hline
    \color{mygreen}{2}& $37.44^\circ$ & 7.2 &$52.14^\circ$ & 2.06 & 2.98\\
    \hline
    \color{blue}{3}& $37.44^\circ$ & 7.6 &$52.71^\circ$ & 2.84 & 3.41\\
    \hline
  \end{tabular}
\caption{\label{tab:neutrino}Numerical values associated with the sample points displayed in Figure \ref{fig:neutrino}.}
\end{table}

Pure Dirac masses have been assumed for the neutrinos to put all the leptons on an equal footing.  Needless to say that the introduction of Majorana masses to implement the seesaw mechanism would imply less stringent constraints on the masses and mixing angles. In particular, we have checked that this is already the case for degenerate Majorana masses.

\section{Taming the strong hierarchy in the up quark family}
As already mentioned, a naive estimate shows that relation \eqref{eq:koide} might be valid for the down quark family but certainly not for the up quark one, because of the large top mass.  However, if a quark-lepton connection exists beyond the Standard Model, $\theta_{13}= 0$ could be a property shared by all elementary fermions. Viewing $U_L^e=\mathbbm{1}$ as a basis fixing choice, one may then write
\begin{equation}
  \label{eq:Uud}
  U_L^u=\Phi(\varphi_u) R_{23}(\theta_t)R_{12}^T(\theta_u)\quad,\quad  U_L^d=\Phi(\varphi_d) R_{23}(\theta_b)R_{12}^T(\theta_d)
\end{equation}
where
\begin{equation}
\Phi(\varphi)\equiv \textrm{diag}(e^{-i\varphi_1},e^{-i\varphi_2},e^{-i\varphi_3})\quad .
\end{equation}
If $U_R=\mathbbm{1}$, the number of arbitrary phases could be reduced ($\varphi_u=-\varphi_d$) by imposing the auxiliary condition $\arg\det(M^u M^d)=0$ from the conspicuous time-reversal invariance of the strong interactions. But anyhow, the pseudo-masses defined in \eqref{eq:WBmass} only depend on the small rotation angles, not on the phases.  So, let us try to extract these angles from the data.  

Suitable rephasing of the quark fields leads to the CKM mixing matrix
\begin{equation}
\label{eq:CKM}
 V_{CKM}\equiv U_L^{u^\dag} U_L^d = R_{12}(\theta_u) \ \mathrm{diag}(e^{-i\varphi},1,1)\  R_{23}(\theta)\ R_{12}^T(\theta_d)
\end{equation}
where $\theta\equiv \theta_b-\theta_t$. Here, the $CP$-violating phase $\varphi$ is only linked to the first and second families. This parametrization coincides with the one convincingly advocated in \cite{Fritzsch:1997st, Fritzsch:1997fw} on the basis of the hierarchical structure of the quark mass spectrum. We find it quite interesting to reach the same description of the CKM mixing matrix from two different approaches.  The parameters can be computed at the $3 \sigma$ level using the latest experimental data \cite{Charles:2004jd} for the absolute values of the $V_{CKM}$ elements
\begin{equation}
\begin{split}
\theta &= (2.37 \pm 0.13)^\circ\\
\theta_u &= (5.37 \pm 1.09)^\circ\\
\theta_d &= (11.53 \pm 2.75)^\circ\\
\varphi &= (95.2 \pm 16.8)^\circ\quad.
\end{split}
\end{equation}
So, we are just left with the freedom on one mixing angle (say $\theta_t$) to test the pseudo-mass relation \eqref{eq:new} for the up and down quark families.  Contrary to the leptons, the quark masses are not directly measurable quantities.  If the quark masses are chosen at a common energy scale $M_Z$ (see Table~\ref{fig:masses}), both relations can be reasonably satisfied (i.e. $\tilde q^u = \tilde q^d \simeq 0.71$) for $\theta_t = -2.21^\circ$.  Conversely,  imposing $\tilde q^{u,d} = \tfrac{2}{3}$ gives $\theta_t = -3.57^\circ$ together with $m_s(M_Z) = 136\ \pm\ 15\text{ MeV}$,
if the rather stable ratios of the light quarks  are used.
  
These theoretical results are quite encouraging once one realizes that relation \eqref{eq:new} is energy scale dependent. The heavy quark mixing ($\theta$) and masses ($m_{t,b}$) are indeed subject to renormalization-group effects. In particular, the running of masses beyond the leading log approximation in QCD flattens the hierarchy such that $\tilde q$ decreases with increasing energy. On the other hand, the corresponding QED effect on relation \eqref{eq:koide} is negligible.

\begin{table}
\begin{tabular}{|r@{ = }l||r@{ = }l|}
\hline
$m_u(M_Z)$ & $1.7 \pm 0.4 \text{ MeV}$ & $m_c(M_Z)$ &  $0.62 \pm 0.03 \text{ GeV}$ \\
$m_d(M_Z)$ & $3.0 \pm 0.6 \text{ MeV}$ & $m_b(M_Z)$ &  $2.87 \pm 0.03 \text{ GeV}$ \\
$m_s(M_Z)$ & $54 \pm 11 \text{ MeV}$ & $m_t(M_Z)$ &  $171  \pm 3 \text{ GeV}$ \\
\hline
\end{tabular}
\caption{\label{fig:masses}Quark masses at the Z mass scale \cite{Jamin}}
\end{table}

\section{Towards a universal mass relation}
There have been few attempts to explain the factor $q=\tfrac{2}{3}$ appearing in relation \eqref{eq:koide} for the charged leptons, but none of them are convincing so far \cite{Koide:2005}.  Here, we have argued that relation \eqref{eq:new} might give us a clue for the understanding of the lepton \emph{and} quark mass spectrum.  

From a theoretical point of view, the numerical factor $\tilde q=\tfrac{2}{3}$ appearing in this universal mass relation should be explained on the basis of symmetry arguments. It is well-known \cite{Georgi:1979} that mass ratios like $\frac{m_b}{m_\tau} = 1$ or $\frac{m_s}{m_\mu} = \frac{1}{N_c}$ ($N_c$ being the number of colours) naturally result from quark-lepton grand unification at high scale.  Similarly, one may hope that $\tilde{q}=\frac{2}{N_f}$ ($N_f$ being the number of families) eventually arises from flavour symmetries broken above the Fermi scale. In fact, the factorized CKM mixing matrix given in \eqref{eq:CKM} is well-suited for specific models of quark mass matrices. Consequently, textures corresponding to the boundary value $\tilde{q}=1$ or $\tilde{q}=\frac{1}{N_f}$ could constitute a good starting point for flavour model building.

With regard to experimental constraints, our main predictions concern neutrino physics with $m_1=(3\pm 1) \, 10^{-2} \, \mathrm{eV}$ and $\theta_{\oplus}>\pi/4$.

\begin{acknowledgements}
This work was supported by the Belgian Federal Office for Scientific, Technical and Cultural Affairs through the Interuniversity Attraction Pole P5/27.
\end{acknowledgements}

\bibliography{biblio}

\end{document}